\def\be{\begin{equation}}
\def\ee{\end{equation}}

\documentclass[12pt,a4paper,epsfig]{iopart}
\usepackage{epsfig}
\newcommand{\labs} {\left\vert}
\newcommand{\rabs} {\right\vert}

\newcommand{\lrb} {\left(}
\newcommand{\rrb} {\right)}
\newcommand{\lcb} {\left\{}
\newcommand{\rcb} {\right\}}
\newcommand{\lab} {\left\langle}
\newcommand{\rab} {\right\rangle}

\usepackage{color}
\begin{document}
\title[Theoretical studies of spin-dependent electrical transport through carbon nanotbes]
      {Theoretical studies of spin-dependent electrical transport through carbon nanotbes}
\author{\bf S. Krompiewski}
\address{Institute of Molecular Physics, Polish Academy of
Sciences, ul.~M.~Smoluchowskiego 17, 60179 Pozna{\'n}, Poland}
%\author{\bf A.K.B}
%\address{}
\begin{abstract}
Spin-dependent coherent quantum transport through carbon nanotubes
(CNT) is studied theoretically within a tight-binding model and
the Green's function partitioning technique.
%Interfaces of
End-contacted metal/nanotube/metal systems are modelled and next
studied in the magnetic context, i.e. either with ferromagnetic
electrodes or at external magnetic fields. The former case shows
that quite a substantial giant magnetoresistance (GMR) effect
occurs ($\pm 20\%$) for disorder-free CNTs. Anderson-disorder
averaged GMR, in turn, is positive and reduced down to several
percent in the vicinity of the charge neutrality point. At
parallel magnetic fields, characteristic Aharonov-Bohm-type
oscillations are revealed with pronounced features due to a
combined effect of: length-to-perimeter ratio, unintentional
electrode-induced doping, Zeeman splitting, and energy-level
broadening. In particular, a CNT is predicted to lose its ability
to serve as a magneto-electrical switch when its length and
perimeter become comparable. In case of perpendicular geometry,
there are conductance oscillations approaching asymptotically the
upper theoretical limit to the conductance, $4 e^2/h$. Moreover in
the ballistic transport regime, initially the conductance
increases only slightly with the magnetic field  or remains nearly
constant because spin up- and spin down-contributions to the total
magnetoresistance partially compensate each other.
\end{abstract}
%\jl{3}
%\ead{stefan@ifmpan.poznan.pl}
% Nanotubes, GMR, Ballistic magnetoresistance
\maketitle
\section{Introduction}
 Recently a new field of electronics
called spin electronics or just spintronics has been rapidly
developing. In comparison with conventional electronics the most
obvious advantage the spintronics offers is a possibility of
controlling current flowing through mesoscopic or nanoscopic
systems with magnetic field, by making use of the spin degree of
freedom of electron. The well-known example illustrating this, is
the famous giant magnetoresistance (GMR) effect, initially
discovered in all-metal multilayers \cite{Baibich}, and next also
in all-semiconductor tunnel junctions \cite{Tanaka}, and molecular
systems \cite{Xiong}. Here the main interest will be focused on a
very peculiar class of the latter - namely carbon nanotubes (CNT)
\cite{Tsuka}. The GMR effect can only be observed when some parts
of a circuit (usually electrodes) are ferromagnetic, then magnetic
field polarizes them making thereby the electrical transport
spin-dependent (see e.g \cite{Zutic, Sanvito} for excellent
state-of-the-art reviews). Quite interesting quantum effects do
also appear in non-magnetic systems in the presence of external
magnetic field. The carbon nanotube are quite exceptional in this
context, as they may undergo magnetic field-induced major changes
in their energy band structure, resulting even in turning a
metallic CNT into a semiconductor or vice versa. This phenomenon,
predicted theoretically \cite{AA, Saito}, has been recently
confirmed experimentally, so it is clear now that axial magnetic
field, if strong enough, may lead to successive opening and
closing of an energy gap between the conduction and valence bands.
Some of these phenomena occurring in CNTs, e.g the Aharonov-Bohm
(AB) effect, as well as conductance oscillations in perpendicular
magnetic field due to the Landau-like quantization will be also
addressed in the sequel. It is noteworthy that very much like the
GMR effect, believed to have practical applications in CNT-based
spintronic devices, also the AB effect in nanotubes might be used
to construct magneto-optical or magneto-electrical switches. One
of the objectives of this study is to elucidate the importance of
size effects directly related to miniaturization problems and the
possibility of using tiny CNTs as elements of functional devices
and interconnects in nanoelectronics. Ultra short CNTs, with
comparable circumferences and lengths, have not yet been
experimentally studied as far as their transport properties are
concerned, in spite of the fact that some other ultra short
molecular systems, including fullerene, were successfully
electrically contacted and measured (\cite{Smit},
\cite{Pasupathy}). The shortest carbon nanotube segments operated
with so far have been about 20 nm long \cite{Postma}. Such
segments may be created with an AFM tip in a double-buckle
nanotube form.

The paper is organized as follows: In Sec.~2 the computational
method based on a tight-binding model is shortly outlined. Sec.~3
is devoted to the GMR effect, whereas in the subsequent section
(Sec.~4) the effect of magnetic field on electrical transport is
discussed. Finally main results of the paper are summarized.

\section{Methodology and Modelling}

\begin{figure}[b]
\centering \epsfxsize=4in \epsfysize=4in
\rotatebox{0}{\epsfbox{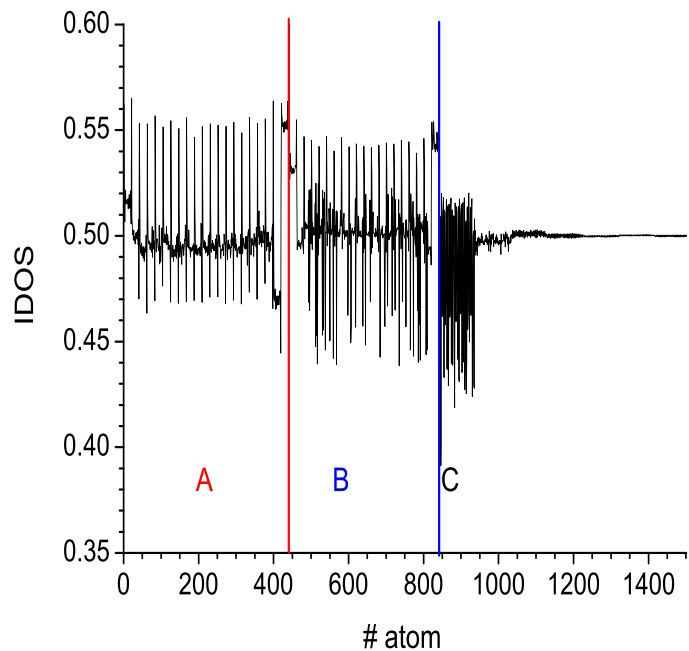}} \caption{Atom projected
integrated density of states per spin (IDOS) at the metal
electrode (fcc-(1,1,1)) and the (24,24)-SWCNT. Labels A, B denote
the two electrode interface-planes (with $21^2$ and $20^2$ atoms,
respectively), whereas C denotes the carbon atom region (96 atoms
at the interface unit cell). Note large fluctuations in the IDOS
at the metal/SWCNT interface.} \label{NOS}
\end{figure}

 We use a tight-binding model for non-interacting $\pi$-electrons
 within nanotubes and $s$-electrons in each of the two
 metal-electrodes. In the case of ferromagnetic electrodes
 the s-electron spin-split band is supposed to mimic d bands of real
 transition metals.
 The total Hamiltonian reads
%\begin{equation} \label{H}
\be \label{H} H = \sum \limits_{ i ,j, \sigma } t_{i,j}\labs i,
\sigma \rab \lab \sigma, j \rabs +\sum \limits_{i, \sigma}
\epsilon_{i, \sigma} \labs i, \sigma \rab \lab \sigma, i \rabs,
%\end{equation}
\ee where i and j run over the whole device (i.e the CNT itself
and the electrodes), $\sigma$ is the spin index, and $t_{i,j}$ and
$\epsilon_{i,\sigma}$ stand for the hopping integrals and the
on-site potentials, respectively. This is an independent-electron
model, applicable in the case of high-transparency contacts, and a
relatively long mean free-path, i.e. when electrical transport can
be regarded as quasi-ballistic (\cite{Liang} - \cite{Li}). \\
\indent Carbon nanotubes have been simulated by means of codes
based on those of Ref.~\cite{Saito} (see Appendix therein). The
model electrodes are fcc(1,1,1) leads infinite in all the 3
directions contacted to CNTs via a constriction composed of 2
finite atomic planes of each electrode. So the whole system
(device) considered here is end-contacted, in contrast to other
wide-spread geometries, like those of side-contact- and
embedded-contact-schemes. The device has been relaxed in order to
find energetically favorable positions of interface atoms.
 The interface metal
atoms have been modelled  by spheres $d_M$ =2.51 $\AA$ in diameter
(this figure roughly fits to the most common metal electrodes).
Carbon atoms in turn, have been represented by smaller spheres
with diameter equal to $d_C =1.421$ $\AA$. Using simple
geometrical arguments it is assumed that interface CNT atoms tend
to be as close as possible to their neighboring metal atoms (big
spheres). So the Lennard-Jones potential (any other would also
work) is simply $(\sigma/r)^{12}-(\sigma/r)^6$ with a parameter
$\sigma = (d_M+d_C)/2$ and an irrelevant pre-factor set to 1.
Hereafter, the CNT together with the aforementioned extra atomic
planes will be referred to as the extended molecule. The way the
on-site ($\epsilon_i$) potentials are determined is as follows.
%Fig.~\ref{scheme}.
%%%%%%%%%%%
First the $\epsilon_i$-s are set equal to those of separated
components of the device, i.e. for CNTs $\epsilon_i=0$ (charge
neutrality point), and for electrodes -- to a certain
nonzero-value chosen so as to give a required number of electrons
per atom. In the ferromagnetic case spin-dependent on-site
potentials must additionally give a required magnetic moment per
atom. This has been implemented by making a rigid band (Stoner)
splitting in the metal electrodes. The on-site potentials
referring to intrinsically non-magnetic carbon atoms have not been
split. Of course when the electrodes are magnetic (Sec. 3),
self-consistent calculations result in inducing slight magnetic
moments at interface carbon atoms. In the present model
calculations an average number of electrons per metal atom is
equal to n=1 and its spin polarization, defined as $P=100
(n_\uparrow-n_\downarrow)/(n_\uparrow+n_\downarrow)$, quals 50\%.
Second, an extra self-consistent potential has been added to
diagonal matrix elements of the Hamiltonian referring to the
extended molecule. At the first step of the self-consistency loop
a self-consistent potential is set 0, whereas during the next
steps it is modified so as to give no global charge excess on the
extended molecule, with the total charge given by the trace of the
density matrix $\hat n$ (Eq. 3, see below). This simple procedure
makes it possible to line up the CNT and metal electron energy
bands, and fix the position of the Fermi level, in particular. As
an example, Fig.~\ref{NOS} shows the number of electrons per atom
in a device composed of infinitely large electrodes (in the x, y,
z directions, not shown) end-contacted via finite two-plane necks
(A, B) to a (24,~24)-CNT (region C). Figure~\ref{NOS} is a 1-d
visualization of a 3-d system with atoms aligned as in a chain, so
atoms labelled with consecutive numbers are usually not nearest
neighbors. Abrupt charge changes visible in this figure in the A
and B regions are due to edge atoms with reduced neighborhood,
which do not contribute effectively to electron scattering.
%%%%%%%%%%%%%%%%%%%%%%%%%%%%%%%%%%%%%%%%%%%%%%%%%%%%%%%%%%%%%%%%%%%%%%%%%%%%%
%%%%%%%%%%%%%%%%%%%%%%%%%%%%%%%%%%%%%%%%%%%%%%%%%%%%%%%%%%%%%%%%%%%%%%%%%%%%%%
It is instructive to note that our simple model gives a
qualitatively correct physical picture with strong oscillations in
the number of electrons per atom in the vicinity of the interface
which next get weakened and die down. It should be however made
clear that a more direct way to align the Fermi levels, while
dealing with real electrodes, would consist in taking into account
work functions (WF) of involved materials (carbon vs. metal). In
fact the work function of a CNT is usually supposed to be roughly
equal to that of the graphite i.e. 4.5 eV (as a matter of fact it
is however chirality- and diameter-dependent), whereas the most
frequently used metal electrodes may have either a bigger WF than
that (Au, Pd and Fe, Co, Ni 3d-transition metals) or a smaller one
(e.g. Ti) \cite{Xue}. The former typically results in transferring
charge from the CNT to the electrode, whereas the latter -- the
other way round. As it follows from Fig.~\ref{NOS}, the model
under consideration here mimics the former case, with a noticeable
electron deficit at the CNT interface.

The Green function of the extended molecule ($G_C$), as well as
the density matrix ($\hat n$) and the conductance per spin (G) are
defined in a standard way as
\begin{equation} \label{eq:G}
G_C =\lrb  E - H_{\rm C}-\Sigma_{\rm L}-\Sigma_{\rm R}\rrb ^{-1},
\end{equation}
\begin{equation} \label{n}
\hat n= \frac{1}{2 \pi} \int dE \, G_C \lrb f_{\rm L} \Gamma_{\rm
L}+f_{\rm R} \Gamma_{\rm R} \rrb {G_C}^\dagger,
\end{equation}
\begin{equation} \label{G}
G= \frac{e^2}{h} \textrm{Tr}\, \lcb \Gamma_{\rm L} \;G_C
\;\Gamma_{\rm R} \;{G_C}^\dagger \rcb,
\end{equation}
where the respective subscripts C, L and R refer to the extended
molecule (central part) and the left and right electrodes. In
contrast to rather common wide-band type approximations, the
present method is based on self-energies ($\Sigma$ matrices) which
are energy-dependent and have been expressed in terms of
recursively computed surface Green functions $g(E)$ of infinite
leads (electrodes) as well as the extended-molecule/electrode
coupling matrices $V$:
\begin{equation} \label{sigma}
 \Sigma_\alpha=V g_\alpha V^\dagger.
\end{equation}

The other quantities in Eq.~(\ref{n}) are the line-width (or
broadening) matrices $ \Gamma_\alpha = i ( \Sigma_\alpha
-\Sigma_\alpha^\dagger ) $ and the Fermi-Dirac distribution
functions $f_\alpha$, with $\alpha=L, R$.

\section{Giant magnetoresistance in ferromagnetically contacted
carbon nanotubes}

Since the pioneering paper on the GMR effect in CNT-based devices
was published \cite{Tsuka}, there has been a lot of interest in
electrical transport through ferromagnetically contacted
nanotubes. It is now clear that the spin diffusion length in
nanotubes is quite long, and often exceeds a separation length
between source and drain electrodes, which is routinely made as
short as a few hundred nm nowadays. This means that the CNT may be
quite attractive for spintronic application either as
interconnects or active elements. So far most of GMR experiments
on CNTs have been performed on relatively high resistive devices
\cite{Tsuka, Zhao, Kim}, corresponding to the tunnelling regime
with resistance (R) of the order of $M\Omega$. Notably, quite
recently the Basel group \cite{Sahoo, preprint} succeeded in
fabricating devices whose resistance was substantially decreased
down to the order of quantum resistance ($h/e^2=26.8 \; k\Omega$).
It is well-known from independent studies of the Stanford
\cite{Javey} and Basel \cite{Babic} groups that Pd makes excellent
contacts with CNTs. The above-mentioned new highly transparent
ferromagnetic contacts do also contain palladium
(Pd$_{0.3}$Ni$_{0.7}$ alloy). In what follows, we calculate the
giant magnetoresistance, defined (in a "pessimistic way") as
\begin{equation} \label{GMR}
GMR= ( G_{\uparrow,\uparrow}-G_{\uparrow,\downarrow})/
G_{\uparrow,\uparrow},
\end{equation}
where the arrows denote the aligned and anti-aligned
magnetizations of the metal electrodes, with G defined by
Eq.~\ref{G}. A system under consideration is an (8,8)-SWCNT both
disorder-free and disordered. The Anderson type disorder is
introduced by randomizing carbon on-site potentials and letting
them vary in an interval [-W/2, W/2], where the disorder parameter
W is set to 0.2t. This simple approach is intended to test whether
or not in the case of GMR, a multiwall CNT can effectively be
treated as equivalent to just its outer shall subjected to some
perturbation from the inner shells.
%%%%%%%%%%%%%%%%%%%%%%%%%%%%%%%%%%%%%%%%%%%%%%%%%%%%%%%%%%%%%%%%%%%%%%%%%%%%%
\begin{figure}[b]
\centering \epsfxsize=3.1in \epsfysize=3.1in
\rotatebox{0}{\epsfbox{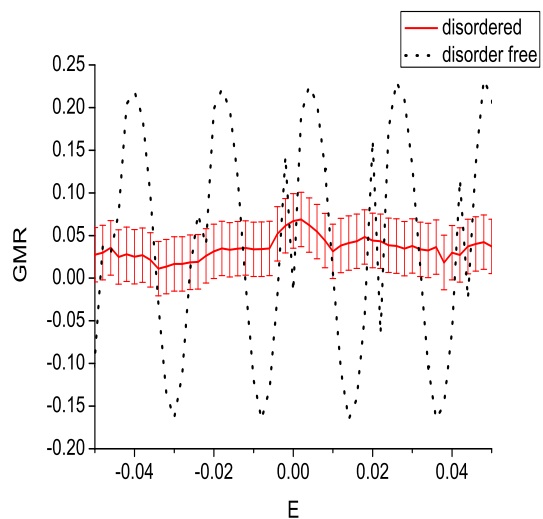}} \caption{GMR of an (8,8)
single-wall CNT \emph{ca} 30 nm long, for the disorder-free
(dotted curve) and disordered (solid line) cases. The latter is
computed from the disorder-averaged conductances (over 100 random
configurations), standard-deviation error bars are also shown. The
presented energy range is close to the charge neutrality point
(E=0), with the hopping integral chosen as the energy unit
($\left| t \right|$=2.7 eV).} \label{GMR88}
\end{figure}
%%%%%%%%%%%%%%%%%%%%%%%%%%%%%%%%%%%%%%%%%%%%%%%%%%%%%%%%%%%%%%%%%%%%%%%%%%%%%%
This hypothesis is generally believed to be acceptable by both
theorists \cite{Egger, Triozon} and experimentalists
\cite{Bachtold, Stojetz}. Figure~\ref{GMR88} shows how the GMR
effect gets modified under the influence of Anderson-disorder. The
obvious observations are that disorder destroys periodicity of GMR
and decreases its absolute value from roughly 20$\%$
(\cite{Mehrez},\cite{ustron}) down to a few per cent
simultaneously reducing tendency of GMR to become negative. All
these features are in accord with experimental data on GMR in
multiwall carbon nanotubes (MWCNTs) with transparent ferromagnetic
contacts \cite{Sahoo, preprint}. The most notable effect of the
disorder is removing of the characteristic energy scale ($\Delta
E$) in the conductance spectrum. In the quasi-ballistic regime
$\Delta E$ is related with energy-level spacings, and scales
inversely proportional to the CNT length (see e.g. \cite{KNC}).
For less transparent contacts $\Delta E$ must be completed with a
charging energy, i.e. the so-called addition energy is to replace
the $\Delta E$ parameter. The double-peak structure visible in
Fig.~\ref{GMR88} (dotted curve) results from lifting of the spin
degeneracy by the ferromagnetic electrodes. The splitting effect
is however too subtle to be visible in the presence of disorder
(solid curve).

\section{Nanotubes in parallel and perpendicular magnetic fields}

 As regards a parallel (axial)
field configuration, it was predicted theoretically that magnetic
field can drastically change electronic band structure of carbon
nanotubes \cite{AA, Saito}. This may lead to opening of the energy
gap in an initially metallic CNT and turn it into a semiconducting
one. The same applies to a semiconducting CNT, which may reveal a
more and more reduced gap with increasing magnetic field, and
eventually become metallic. Recently, two important experimental
papers have been published, that fully confirm aforementioned
scenarios. The first one \cite{Coskun} visualizes in a direct way
magnetic flux modulation of the energy gap in high-diameter
MWCNTs, with an outer radius of about 15 nm which makes it
possible to get the magnetic flux quantum $\Phi_0=h/e$ at
accessible fields of $B \sim 6$T. The AB effect can also be
detected in much thinner single-wall carbon nanotubes, typically a
few nm in diameter, although in this case obviously just a small
part of the AB phase comes into play, nevertheless it is
detectable by means of magneto-optical techniques under pulsed
magnetic fields \cite{Zaric}. In this section we generalize the
hitherto existing theoretical studies restricted to free CNTs, by
taking into account the presence of electrodes and thereby on the
one hand, the accompanying charge transfer, and on the other hand
- the energy level broadening which becomes quite crucial for
ultra short CNTs with transparent contacts. Incidentally, the
charge transfer problem is seldom addressed in theoretical papers
devoted to magnetoresistance problems, as it requires integration
over all occupied states and makes calculations extremely
computer-time consuming and expensive. Therefore theoreticians
often skip the charge transfer issue and restrict themselves only
to conductance calculations, which may be carried out just at the
Fermi energy (or within a narrow "transport energy window"). The
present studies go beyond this limitation.

 We use the well-known Peierls substitution \cite{Saito}
%%%%
%%%%
\begin{equation} \label{Peierls}
t \longrightarrow t \;\exp[i 2 \pi \varphi /\Phi_0].
\end{equation}
 The
factor $\varphi$ is defined in terms of a magnetic flux $\Phi=B
\pi (\frac{C_h}{2 \pi})^2$ penetrating a cross-section of the
(m,n) CNT with perimeter $C_h=a \sqrt{n^2+m^2+mn}$, where B stands
for a uniform static magnetic field and the graphene lattice
constant $a=2.49 \AA$ is a length unit of the theory
\begin{equation} \label{zeta}
\varphi=\cases{ \frac{\Delta x}{C_h}\Phi, &
$\!\!\!\!\!\!\!\!\!\!\!\!\!$ parallel field \cr \left(\frac{C_h}{2
\pi}\right)^2 \frac{B \Delta y}{\Delta x} \left[\cos \frac{2 \pi
x}{C_h}-\cos \frac{2 \pi (x+ \Delta x)}{C_h}\right], &
$\!\!\!\!\!\!\!\!\!\!\!\!\!$ perpendicular field. \cr}
\end{equation}
%%%%%%%%%%%%%%%%%%%%%%%%%%%%%%%%%%%%%%%%%%%%%%%%%%%%%%%%%%%%%%%%%%%%%%%%%%%%%%
\begin{figure}[b]
\centering \epsfxsize=3.5in \epsfysize=3.5in
\rotatebox{0}{\epsfbox{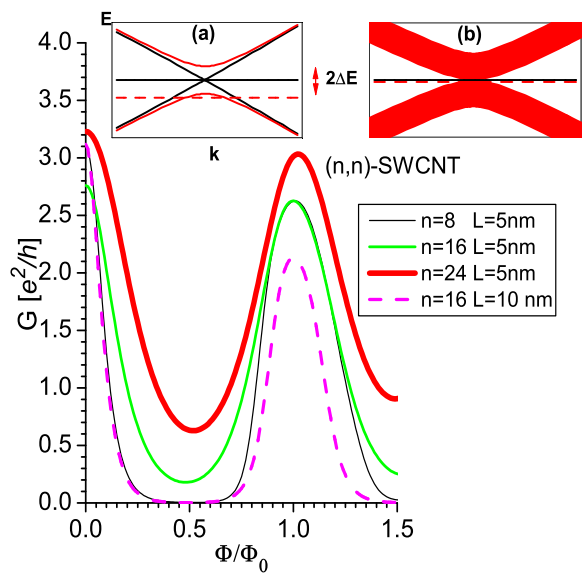}} \caption{Magnetoconductance of
small armchair SWCNTs in a parallel field \emph{vs} magnetic flux.
The inset (a) shows that when the doping is high the Fermi energy
$E_F$ (dashed horizontal line) might lie outside the maximum
magnetic field-induced energy gap, $2\Delta E=\pi \left|t \right|
/n-g \mu_B B$, and the conductance never vanishes. In fact the
same effect occurs for a much smaller $E_F$-shift provided the
energy-level widths, originating from a strong CNT-electrode
coupling are broad enough (Inset (b)).} \label{AB-arm}
\end{figure}
%%%%%%%%%%%%%%%%%%%%%%%%%%%%%%%%%%%%%%%%%%%%%%%%%%%%%%%%%%%%%%%%%%%%%%%%%%%%%%
Circumferential and axial coordinates of carbon atoms are $x$,
$x+\Delta x$, and $y$, $y+\Delta y$, respectively. It is
noteworthy that within the present approach magnetic field enters
not only off-diagonal block-elements of the Hamiltonian (via the
Peierls substitution) but also the diagonal ones through the
Zeeman splitting, $\pm \frac{1}{2} g \mu_B B$. The relevance of
the Zeeman splitting in the AB effect is often underestimated, but
theoretical results of \cite{Jiang} as well as measurements
presented in \cite{Coskun} show that it can lead to fairly
pronounced effects. In fact the Zeeman splitting is taken into
account here not only within the CNT but also in metal leads,
since in the case of small-size devices there is no way to apply
magnetic field locally without interfering the electrodes.
 The Peierls
substitution has not however been applied to contacts. The reason
is that it could be
  hardly implemented in the present recursive calculation scheme of infinite
  electrodes. Fundamentally different geometries of the
  SWCNTs and contacts, as well as strong screening effects within the (metal)
  contacts justify additionally this approximation.
Figure~\ref{AB-arm} shows Aharonov Bohm oscillations in selected
(n,n)-SWCNTs contacted to the model paramagnetic fcc-(1,1,1)
electrodes.
%%%%%%%%%%%%%%%%%%%%%%%%%%%%%%%%%%%%%%%%%%%%%%%%%%%%%%%%%%%%%%%%%%%%%%%%%%%%%%
\begin{figure}[b]
\centering \epsfxsize=3.5in \epsfysize=3.5in
\rotatebox{0}{\epsfbox{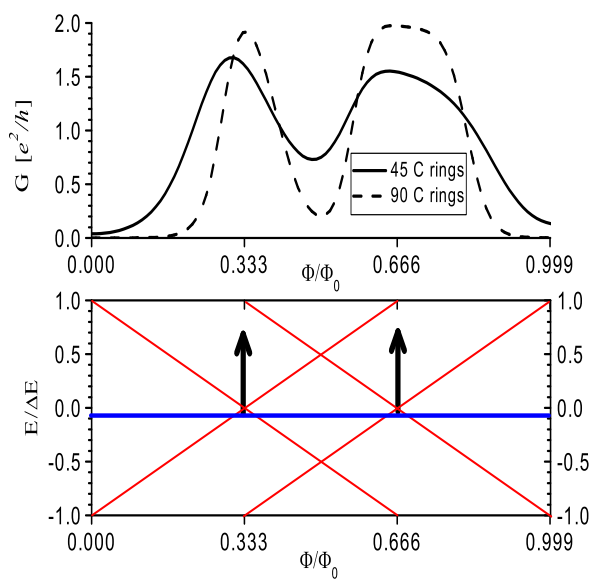}} \caption{Top panel: Conductance
of a (14,0)-SWCNT \emph{ca} 5 nm (solid line) and 10 nm (dashed
line) long vs. dimensionless magnetic flux. Bottom panel: Sketch
for closing and opening of the energy gap (2 $\Delta E$),
according to the AA theory. Vertical arrows indicate anticipated
conductance peaks. The comparison of the panels shows clearly that
energy levels are considerably broadened in electrically contacted
nanotubes.} \label{AB140}
\end{figure}
It is shown that conductance is a quasi-periodic function of
$\Phi$ with a period equal to $\Phi_0$. Remarkably, in general the
conductance does not drop to zero at $\Phi=\Phi_0/2$. This might
have been due to the fact that there is some charge transfer in
the system which locates the Fermi energy beyond a magnetic
field-induced energy gap (see the inset \emph{a}). In fact however
it results from the calculations that the accompanying energy
shift is by far smaller than half the magnetic field-induced
energy-gap. To understand the situation it must be realized that
in devices with transparent contacts energy levels are quite
broad. A zero-field energy level spacing $\delta E$ is strongly
dependent on nanotube axial length (L). In the ballistic regime
$\delta E=\pi \sqrt 3 \left|t \right| a /(2L)$ and the energy
level broadening also scales as $1/L$ \cite{KNC}. Thus for
small-size nanotubes studied in this section with $L \sim 20a$ --
$40a$, the energy-level width turns out to be comparable with the
energy-gap $\Delta E$. These two energy scales are close to each
other because of comparable magnitudes of the perimeter (sampled
with the axial field) and the length. So a crucial role should be
ascribe to the interplay between the energy-level broadening
effect and the magnetic field-induced energy gap whose maximum
value is proportional to the inverse nanotube radius.
Consequently, as shown in Fig.~\ref{AB-arm} the AB effect is very
strongly dependent on the nanotube perimeter as far as the
collapse of metal-semiconductor switching ability is concerned.
The crossover takes place at the circumference-to-length ratio
close to unity. This is clearly visible for the SWCNTs with n=16
(perimeter $C_h$=7nm), which still have a considerable conductance
at $\Phi/\Phi_0$=0.5 for L=5 nm (middle curve), but not for L=10
nm (dashed curve). It implies that if assumptions of the present
theory were fulfilled then, e.g. a (150,150) CNT would lose its
magneto-electrical switching efficiency for a length below
\emph{ca} 65 nm.

%%%%%%%%%%%%%%%%%%%%%%%%%%%%%%%%%%%%%%%%%%%%%%%%%%%%%%%%%%%%%%%%%%%%%%%%%%%%%%
\begin{figure}[b]
\centering \epsfxsize=4in \epsfysize=4in
\rotatebox{0}{\epsfbox{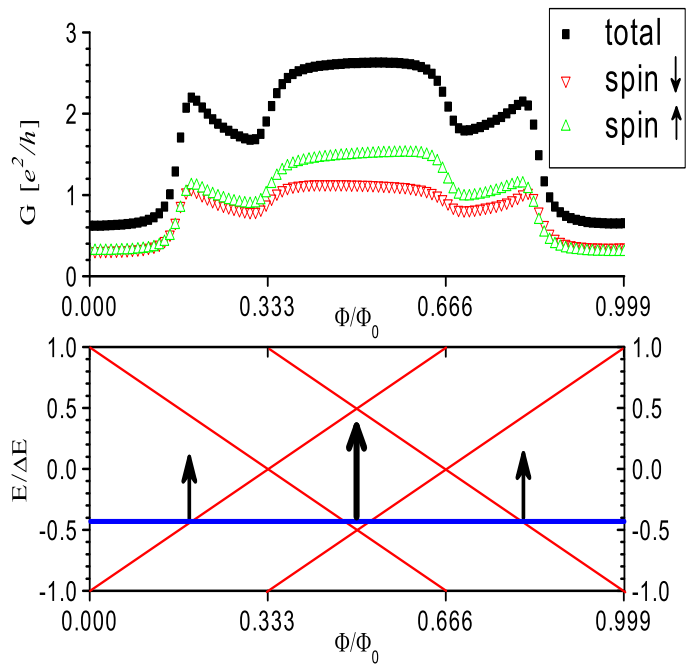}} \caption{Same as
Fig.~\ref{AB140} but for a (28,0)-SWCNT. The lower panel
represents a possible interpretation of the upper one within the
AA theory. In fact however the Fermi energy is always very close
to E=0, but the energy level widths are of the order of $\Delta
E/2$ for ultra short devices (\emph{cf} Inset (b) in
Fig.~\ref{AB-arm}).} \label{AB280}
\end{figure}
%%%%%%%%%%%%%%%%%%%%%%%%%%%%%%%%%%%%%%%%%%%%%%%%%%%%%%%%%%%%%%%%%%%%%%%%%%%%%%

If the CNTs in question are nominally semiconducting zigzag-type
ones, then again a quasi-period of the conduction oscillations is
$\Phi_0$, but depending on the degree of doping, the nanotube
perimeter (which determines the number of sub-bands and the
maximum energy-gap, $2 \Delta E = 2 \pi \left|t\right|/(\sqrt 3 \;
n)$) as well as the length (responsible for the number of energy
levels within a sub-band), one gets 2- and 3-peak conductance
curves. In principle, the peak positions might be explained in
terms of the Ajiki-Ando theory \cite{AA} and the aforementioned
arguments about the interplay between the doping, the energy
gap-width and the energy-level broadening. It results from
Fig.~\ref{AB140} that deviations from the AA theory are large for
short-length CNTs, and they manifest themselves by broadening of
conductance peaks and disappearance of a conductance gap at
$\Phi/\Phi_0=0.5$. Incidentally a noticeable asymmetry in the
conductance peak-widths in Fig.~\ref{AB140} results from the
Zeeman splitting which increases with the magnetic field. A
pictorial, rather crude explanation of the results in terms of the
AA theory is sketched in the bottom panels of
Figs.~\ref{AB140}~and~\ref{AB280}. The slanting lines over there
represent the lowest energy level evolution as a function of
magnetic flux (no Zeeman splitting is shown for clarity). Ideally
the CNTs remain semiconducting except at the crossing points,
where the energy gap gets closed. Indeed some traces of such a
behavior are really displayed by these two figures. However the
$E_F$-shift, as depicted in Fig.~\ref{AB280} (bottom horizontal
line) is unrealistically large and inconsistent with the numerical
results. As a matter of fact the actual $E_F$-shift is by more
than one order of magnitude smaller than that (about 20 times),
nevertheless the energy gaps get closed due to exceptionally broad
energy level-widths occurring in these small-size devices having
perimeters comparable with a length. Actually the AA theory in its
conventional version applies to perfect, infinite CNTs of
arbitrary chirality. It is however restricted to a low energy
region in the energy spectrum ("light-cone" approximation) and
disregards external contacts and any possible charging effects.
Nevertheless our results show that there are reminiscences of
major AA features even under more complicated conditions than
those originally assumed. It is interesting to note that the
conductance spectra are spin-dependent due to the Zeeman splitting
(up-triangles and down-triangles in Fig.~\ref{AB280}) for high
enough $\Phi$ ($\sim \Phi_0/6$). This suggests that in the
quasi-ballistic regime a spin polarization of the current flowing
through the CNT appears to be controllable by the magnetic field
strength.

\begin{figure}[t]
\centering \epsfxsize=3.5in \epsfysize=3.5in
\rotatebox{0}{\epsfbox{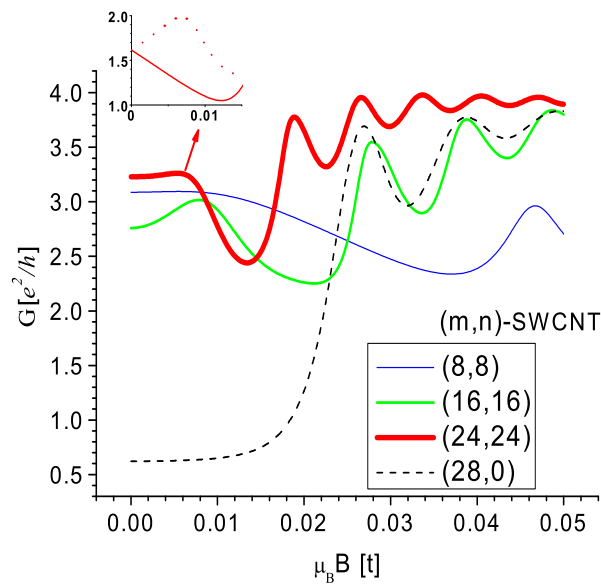}} \caption{Magnetoconductance of
(m,n)-SWCNTs at a perpendicular magnetic field. Note that
quasi-periods of the curves scale inversely proportional with the
CNT diameter, and G tends to $4 e^2/h$. The inset shows
spin-dependent contributions to the total conductance for m=n=24
($G\downarrow$ - dotted line, $G\uparrow$ - solid line).}
\label{Landau}
\end{figure}

\indent Turning now to the perpendicular field, we focused our
attention on the armchair SWCNTs (5 nm long) and the (28,0)-zigzag
one which have been studied above for the axial geometry.
Figure~\ref{Landau} shows magnetic field - conductance dependence
of these nanotubes.

%%%%%%%%%%%%%%%%%%%%%%%%%%%%%%%%%%%%%%%%%%%%%%%%%%%%%%%%%%%%%%%%%%%%%%%%%%%%%%
%%%%%%%%%%%%%%%%%%%%%%%%%%%%%%%%%%%%%%%%%%%%%%%%%%%%%%%%%%%%%%%%%%%%%%%%%%%%%%
Remarkable features of the plot are that the conductances reveal
oscillations with a decreasing amplitude and asymptotically
approach the upper theoretical limit of $4 e^2/h$ at huge fields.
These features can be accounted for by the fact that energy bands
get narrower and narrower as B increases (magnetic length
$l_m=\sqrt{\Phi_0/(2 \pi B) }$ decreases), and eventually form
Landau-like levels, with a decreasing but always finite width. It
was shown in \cite{pss03, JMMM} that interface conditions as well
as charges in a nanotube length can bring an armchair nanotube to
or out of the so-called on-resonance state with a conductance peak
at the Fermi level (\emph{nota bene} the off-resonance state has a
conductance dip at $E_F$). Experimentally such a tuning is
realized more conveniently by means of a gate voltage. As regards
the magnetic field, it spoils the on-resonance conductance and
improves the off-resonance one. It was verified that indeed, there
is such a competition responsible for the behavior of the
magneto-conductance curves depicted in Fig.~\ref{Landau} for
n=m=16 and 24, respectively. The first peak is due to promoting
the Zeeman-split $\downarrow$-spin electron conductance from the
off-resonance type at B=0 to the on-resonance type at the first
peak position, and roughly the opposite process in the case of
$\uparrow$-spin electrons (see the inset). It turns out that the
former process prevails leading to a slightly negative
magnetoresistance (positive magnetoconductance). For n=8 in turn,
initially the spin-dependent conductances give compensating
contributions, resulting in nearly constant net conductance (up to
0.01). It should be stressed that the described magnetoresistance
mechanism develops within the quasi-ballistic regime in
impurity-free systems.
 The effects described above are weaker and of another
field-scale than the well-known interference phenomena, i.e.
weak-localization (WL) and universal quantum fluctuations (UCF)
\cite{Bachtold,Stojetz}, and obviously the physics standing behind
them is completely different. Experimentally the WL contribution
can be identified on the basis of its temperature-dependence, and
the UCF one may be effectively suppressed upon gate
voltage-averaging of the conductance \cite{Stojetz}. So in the
case of the simultaneous presence of the interference phenomena
and the ballistic magnetoresistance described above, the latter
seems separable from the former ones. \\
\indent Another point which needs commenting on is related with
the fact that the present calculations are restricted to thin
nanotubes (with diameters up to 33 $\AA$). In the parallel field
the AB oscillations have a well-defined periodicity ($\Phi_0$)
independently of the thickness, so the present findings could be
applicable in a rather straightforward way to thick MWCNTs.
 In particular, the
Aharonov-Bohm smearing, at the circumference-to-length ratio close
to unity, shown here to be the inherent property of SWCNTs, should
be detectable e.g. in a (150, 150) CNT system (most probably even
in the Coulomb-blockade regime, see \cite{Coskun}) upon improving
contact transparency and introduction of two buckles ca 20 nm away
from each other.

%provided they are within the ballistic regime and have transparent
%contacts.
The situation is more cumbersome for the perpendicular
geometry, where, as seen in Fig.~\ref{Landau}, the G(B) curves
depend critically on the thicknesses involved, and there is no way
to make a pseudo-universal plot (replacing $\mu_B B$ by a magnetic
strength parameter $\nu=R/l_m$ with a radius $R=C_h/(2\pi)$ does
not help at all). However it results from Fig.~\ref{Landau} that
the first magnetoresistance peak roughly appears within the
interval $1<\nu<2$, i.e. whenever the lowest Landau level emerges.
So it is reasonable to anticipate that this trend will also be
present in the case of thick CNTs if the aforemention conditions
(ballisticity and transparency)  are fulfilled. So the expected
critical magnetic field at which magnetoresistance changes its
sign would be, e.g. 6 - 12 T for a (150,150)-CNT with a diameter
$\sim 200 \AA$. Another argument in support to the suggestion that
magneto-conductance in perpendicular magnetic fields is directly
related with the Landau quantization is a finding that
asymptotically for large B, conductance is no longer
chirality-dependent and its behavior is totally determined by a
nanotube diameter. For instance the (16,16)-SWCNT curve of
Fig.~\ref{Landau} closely resembles that for a (28,0)-SWCNT
(dashed curve) with practically the same diameter, for strong
enough $\mu_BB$. This means that magnetic field localizes electron
wave-functions making thereby circumferential boundary conditions
irrelevant (\emph{cf}~\cite{Ando}).

\section{Summary and conclusions}

In this study the attention has been focused on a quasi-ballistic
spin-dependent transport through carbon nanotubes contacted to two
external electrodes. Unlike some other authors, we use neither
seamless electrodes (made also of CNTs) nor the so-called
wide-band approximation. Moreover, while studying magnetic-field
dependence of the conductance we have included the Zeeman
splitting not only in the CNT alone but also in the electrodes. So
we have been in a position to take into account possible charge
transfer and spin-polarization effects, and estimate their
combined impact on the ballistic magnetoconductance.

The relevance of the spin-degree of freedom is demonstrated by
taking into account the following cases: (i) CNT sandwiched
between ferromagnetic contacts, and (ii) nanotubes of various
thickness, length and chirality placed in external magnetic field
oriented either parallel or perpendicular to the tube axis. In the
former, it was shown that the GMR effect in single-wall
disorder-free CNTs may be quite large, exceeding $20\%$ or so. The
GMR is a quasi-periodic function of energy (controllable with a
gate voltage) with a period related to a characteristic
energy-scale for the problem at hand, namely the inter
energy-level distance in the ballistic regime (and by analogy the
addition energy in the Coulomb blockade case). Interestingly, as a
function of energy the GMR may be both positive and negative
(inverse GMR) with roughly the same absolute value. Upon
introducing disorder, the situation substantially changes. Not
only the amplitude of GMR decreases down to several percent but
also its oscillations become aperiodic and the inverse GMR is in
practice no longer present, at least in the vicinity of the charge
neutrality point. Once such features have been recently
established experimentally on MWCNT with transparent ferromagnetic
contacts, it seems that single-wall CNTs with disorder may really
mimic behavior of multiwall CNTs. Putting it in other words, the
present results give credit to the wide spread view that MWCNT
transport properties can be regarded just as due to its outermost
shell disturbed electrostatically by inner shells (assuming
that the latter are out of contact to the leads). \\
\indent As regards the influence of a magnetic field, it has been
shown here that interface conditions, responsible for energy level
widths and the charge transfer, modify the magnetoconductance
spectra for parallel orientation without changing the period of
the Aharonov-Bohm oscillations. In particular, if the charge
transfer shifted $E_F$ happens to touch the closest energy-gap
edge (considerably broadened by the strong coupling to the
electrodes) then the conductance of armchair CNTs remains finite
even at $\Phi=\Phi_0/2$. For the same reason nominally
semiconducting CNTs may have suppressed but finite conductance at
$\Phi=0$ and $\Phi_0$. Moreover the conductance reveals two- and
three-peak structures which have been interpreted here as being
reminiscent features of the Ajiki-Ando theory when either a large
energy-level broadening or a high doping come into play. It is
noteworthy that SWCNTs lose their ability to serve as
magneto-electrical switches
when their length-to-perimeter ratio approaches unity. \\
\indent The magnetoconductance spectra of CNTs at a perpendicular
field, in turn, reveal pronounced oscillations having a
pseudo-period which scales with a CNT radius. It has been found
that in the (quasi) ballistic regime the magnetoresistance changes
its sign at fields related to the onset of the first Landau level.
A clear dying down of magnetoconduction oscillations at the charge
neutrality point (of the extended molecule) with the increasing
magnetic field is due to narrowing of the lowest Landau
level-width.

\ack I thank the EU grant CARDEQ under contract IST-021285-2, and
the Pozna\'n Supercomputing and Networking Center for the
computing time.


\section*{References}
\begin{thebibliography}{99}

\bibitem{Baibich} Baibich M N, Broto J M, Fert A, Nguyen Van Dau F,
Petroff F, Eitenne P, Creuzet G, Friederich A and Chazelas J 1988
\emph{Phys. Rev. Lett.} {\bf 61} 2472
\bibitem{Tanaka} Tanaka M and Higo Y 2001 \emph{Phys. Rev. Lett.} \textbf{87} 026602
\bibitem{Xiong} Xiong Z H, Wu Di, Vardeny Z V and Shi J 2004 \emph{Nature} {\bf 427} 824
\bibitem{Tsuka} Tsukagoshi K, Alphenaar B W and Ago H, 1999 \emph{Nature} \textbf{401} 572
\bibitem{Zutic} Zutic I, Fabian J and Sarma S D 2004 \emph{Rev. Mod. Phys.} \textbf{76} 323
\bibitem{Sanvito} Sanvito S, 2004 \emph{Handbook of Computational Nanotechnology}, American Science Publishers,
California; cond-mat/0505134
\bibitem{AA} Ajiki H and Ando T 1993 \emph{J. Phys. Soc. Jpn.} \textbf{62}
1255
\bibitem{Ando} Ando T 2005 \emph{J. Phys. Soc. Jpn.} \textbf{74}
777
\bibitem{Saito} Saito R, Dresselhaus M S and Dresselhaus G, 1998 \emph{Physical
Properties of Carbon Nanotubes}, Imperial College Press
\bibitem{Smit}  Smit R H M, Noat Y, Untiedt C, Lang N D, van
Hemert M C and van Ruitenbeek J M 2002 \emph{Nature} \textbf{419}
906
\bibitem{Pasupathy} Pasupathy A N, Bialczak R C, Martinek J, Grose J E,
Donev L A K, McEuen P L and Ralph D C 2004 \emph{Science} \textbf{306} 86
\bibitem{Postma} Postma H W Ch, Teepen T, Yao Z, Grifoni M, Dekker C 2001 \emph{Science} \textbf{293} 76
\bibitem{Liang} Liang W J, Bockrath M, Bozovic D,Hafner J H, Tinkham M and Park H 2001 \emph{Nature}
\textbf{411} 665
\bibitem{KMB} Krompiewski S, Martinek J and Barna\'s J 2002 \emph{Phys. Rev. B}
\textbf{66} 073412
\bibitem{Frank} Frank S, Poncharal P, Wang Z L, de Heer W A,
1998 \emph{Science} \textbf{280} 1744
\bibitem{Li} Li H J, Lu W G, Li J J, Bai X D and
Gu C Z 2005 \emph{Phys. Rev. Lett.} \textbf{95} 86601
\bibitem{Mehrez} Mehrez H, Taylor J, Guo H, Wang J, Loland C 2000 \emph{Phys. Rev. Lett.} \textbf{84}
2682
\bibitem{ustron} Krompiewski S 2005 \emph{physica status solidi (b)} {\bf 242} 226
\bibitem{Xue} Xue Y and Rattner M 2003 \emph{Appl. Phys. Lett.} \textbf{83} 2429
\bibitem{Zhao} Zhao N, M{\"o}nch I, M{\"u}hl T and Schneider C M
2002 \emph{Appl. Phys. Lett.} \textbf{80} 3144
\bibitem{Kim} Kim J -R, So H M and Kim J -J 2002 \emph{Phys. Rev. B} \textbf{66} 233401
\bibitem{Sahoo} Sahoo S, Kontas T, Sch\"onenberger C and S\"urgers C 2005 \emph{Appl.
Phys. Lett.} \textbf{86} 112109
\bibitem{preprint} Sahoo S, Kontas T, Furer J, Hoffmann C, Gr\"aber M, Cottet A and
Sch\"onenberger C 2005 \emph{Nature Physics} \textbf{1} 99;
cond-mat/0511078
\bibitem{Javey} Javey A, Guo J, Wang Q, Lundstrom M and Dai H, 2003\emph{ Nature} \textbf{424} 654
\bibitem{Babic} Babi\'c B, Furer J, Iqbal M and Sch\"onenberger C,
2004 in Electronic Properties of Synthetic Nanostructures,
\emph{AIP Conf. Proc.} \textbf{723}, 574; cond-mat/0406626
\bibitem{Egger} Egger R and Gogolin A O 2001 \emph{Phys. Rev. Lett.} \textbf{87} 066401
\bibitem{Triozon} Triozon F, Roche S, Rubio A and Mayou D 2004 \emph{Phys. Rev. B} \textbf{69}, 121410
\bibitem{Bachtold} Bachtold A, Strunk C, Salvetat J -P, Bonard J -M, Forro L,
Nussbaumer T and Sch\"onenberger C 1999 \emph{Nature} \textbf{397}
673
\bibitem{Stojetz} Stojetz B, Miko C, Ferro L. and Strunk C 2005 \emph{Phys Rev. Lett.} \textbf{94} 186802
\bibitem{KNC} Krompiewski S, Nemec N, Cuniberti G 2006 \emph{physica status solidi (b)} {\bf 243} 179
\bibitem{Coskun} Coskun U C, Wei T -C, Vishveshwara S,
Goldbart P M and Bezryadin A 2004 \emph{Science} \textbf{304} 1132
\bibitem{Zaric} Zaric S, Ostolic G N, Kono J, Shaver J,
Moore V C, Strano M S, Hauge R H, Smalley R E and Wei X 2004
\emph{Science} \textbf{304} 1129
\bibitem{Jiang} Jiang J, Dond J and Xing D Y 2000 \emph{Phys Rev. B} \textbf{62}
13209
\bibitem{pss03} Krompiewski S 2003 \emph{physica status solidi (b)} {\bf 196} 29
\bibitem{JMMM} Krompiewski S 2004 \emph{J. Magn. Magn. Mat.} \textbf{272} 1645



\end{thebibliography}
\end{document}